\documentclass[12pt,twoside]{article}   
\usepackage{latexsym} 
\usepackage{mathptm}
\usepackage{amsmath}
\usepackage{amssymb}
\usepackage{graphicx}

\def\beq{\begin{equation}}
\def\eeq{\end{equation}}
\def\beqn{\begin{eqnarray}}
\def\eeqn{\end{eqnarray}}

\begin{document}
 
\title{Comment on arXiv:1104.2019, ``Relative locality and the soccer ball problem,'' by Amelino-Camelia et al.}
\author{Sabine Hossenfelder \thanks{hossi@nordita.org}\\
{\footnotesize{\sl Nordita, Roslagstullsbacken 23, 106 91 Stockholm, Sweden}}}
\date{}
\maketitle

\vspace*{-1cm}

\begin{abstract}
It is explained why the argument in arXiv:1104.2019 does not answer the question how to describe 
multi-particle states in models with a deformed Lorentz-symmetry in momentum space.
\end{abstract}

In \cite{AmelinoCamelia:2011uk}, the authors claim that ``the soccer ball problem does not
occur'' in the theory they are considering. I will show here that this claim is not supported by the argument they have
presented. 

First, let us state the problem. In the theory considered in \cite{AmelinoCamelia:2011uk},
one has a modified addition law for momenta, see Eq. (10) \cite{AmelinoCamelia:2011uk}
\beqn
p^a_\nu\oplus p^b_\nu = p^a_\nu + p^b_\nu - \frac{1}{m_{\rm p}} \Gamma_\nu^{\; [ \alpha\beta ]} 
p^a_\alpha p^b_\beta \label{add} \quad,
\eeqn 
to first order in $1/m_{\rm p}$, hats and tildes omitted, and $m_{\rm p}$ is the Planck mass. 
Small Latin indices label different particles whose momenta are added.
In the theory with a Lorentz-transformation that is non-linear
on momentum space, one cannot take the ordinary sum because that would not
be Lorentz-invariant. This is why one defines this new addition law $\oplus$ which
is designed to maintain Lorentz-invariance.

The soccer-ball problem is that, at least at first sight, if one considers
$N$ states and iterates the above addition law $N$ times, one
generates $N(N-1)/2$ additional terms. Thus, the additional contribution seems to
scales for large $N$ as $N^2/m_{\rm p}$ times the average momentum squared. In particular,
it does not converge for large $N$ and the momentum of a large number of states is
very far off the normal addition of the momenta. 
The problem is that the linear term grows only with $N$, so the nonlinear term 
should not grow faster than that. Note that we were just talking about
a sum, we are not even talking about bound states in particular, just any system with
$N$ constituents.  

One could try to argue that there is a limit
 to $N$ since there are only so-and-so many particles in the whole universe but that
is not the line of argumentation in \cite{AmelinoCamelia:2011uk}. 
In \cite{AmelinoCamelia:2011uk}, the authors have
considered instead the interaction between two objects with a large number of constituents
with the aim of making a statement about the conservation of momentum in that
exchange. That is an interesting question, but does not address the issue of whether
the macroscopic bodies can be meaningfully described to begin with. In section IV C, 
they show that for colinear momenta the addition law is the same as in the normal
case because they have chosen some particular coordinate system in which this is
fulfilled. That however does not describe a
soccer ball but a collection of particles that happen to fly into the same 
direction. The point of actual relevance for the soccer-ball
problem is in section IV D. 

To summarize the argument in the relevant section of the paper, consider a collection of $N$ particles with momenta $p_a$ where $a \in \{1... N\}$.
Now define  $P = N \langle p_\nu \rangle = \sum_a p^a_\nu$, so that $p^a_\nu = \langle p_\nu \rangle + \delta p^a_\nu$ with 
$\sum_a \delta p^a_\nu = 0$. It is not clear here what, physically, $\langle p \rangle$ actually is, since $P$ is not the
$\oplus$ sum of the momenta that is supposedly physically meaningful, but mathematically this is all well-defined and one
can just use it. In section IV D one then finds that the iteration of the addition
law yields
\beqn
\sum_{\oplus} p^a_\nu = \sum p^a_\nu - \frac{1}{m_{\rm p}}\sum_{a<b} \Gamma_\nu^{[\alpha\beta ]}p^a_\alpha p^b_\beta \quad, 
\eeqn 
Note that this sum is not symmetric in $a$ and $b$. It can't because otherwise it would always vanish. 
(That would solve the problem of course, but then there was no modification left.) Now we 
insert the above decomposition $p^a_\nu = \langle p_\nu \rangle + \delta p^a_\nu$, then

\beqn
\sum_{\oplus} p^a_\nu &=& P +  \frac{1}{m_{\rm p}}\Gamma_\nu^{[\alpha\beta ]} \sum_{a<b} (\langle p_\alpha \rangle + \delta p^a_\alpha )(\langle p_\beta \rangle + \delta p^b_\beta) \quad, \nonumber \\
&=& P +  \frac{1}{m_{\rm p}}\Gamma_\nu^{[\alpha\beta ]} \sum_{a<b} ( \langle p_\alpha \rangle \delta p^b_\beta  + \delta p^a_\alpha \langle p_\beta \rangle + \delta p^a_\alpha \delta p^b_\beta) \quad,  \label{here}
\eeqn 
where we have used that the sum for colinear momenta vanishes. In \cite{AmelinoCamelia:2011uk}, the authors
now take the time average over the additional terms. On the average the fluctuations are symmetric, thus
the contraction with the antisymmetric coefficients yields zero. But taking a time average does not help 
to solve the problem since one still needs to know how large the unaveraged term is, so one knows
how large the fluctuations are, to begin with because the expansion
will break down if the sum doesn't converge.

So let us look at these terms in \ref{here}. The first term vanishes because it contains a sum
over all $\delta p$ which is by construction zero. The second term gives
\beqn
\Gamma_\nu^{[\alpha\beta ]} \sum_{a<b} \langle p_\alpha \rangle \delta p^b_\beta  = 
\Gamma_\nu^{[\alpha\beta ]} \langle p_\beta \rangle \sum_{b=1}^N \sum_{a=1}^b \delta p^a_\alpha  \quad.
\eeqn 
Now for $\alpha \neq 0$, one may expect the $\delta p^a_\alpha$ to be of the order $T$, where $T$ is the 
temperature of the soccer ball. One can thus estimate the
sum over $a$ with a random walk in one dimension, so it should be of order $\sqrt{b}$. That is
not exactly correct as that it does not take into account that the sum goes to to zero for $a=N$, but it
gives an impression about the scaling. (One may imagine taking apart the sum into one for $a\leq N/2$ 
and one for $a>N/2$ and applying the argument separately.) For large $N$, the second sum should 
then have a leading term like $N^{3/2}$, so for $\alpha \neq \beta \wedge \alpha,\beta \neq 0$ we get
\beqn
\Gamma_\nu^{[\alpha\beta ]} \sum_{a<b} \langle p_\alpha \rangle \delta p^b_\beta  \approx
\Gamma_\nu^{[\alpha\beta ]} \langle p_\alpha \rangle N^{3/2} T = \Gamma_\nu^{[\alpha\beta ]} P \sqrt{N} T .
\eeqn 
Interestingly, one finds that this term does not grow as quickly as expected, but it still grows 
relative to the linear term. Assuming that the $\Gamma$'s are of order one, one gets a problem with that 
term when $T \sqrt{N} \approx m_{\rm p}$. To put in some numbers, a neutron star has an average temperature of 
some hundred MeV, and contains about $10^{56}$ particles, so $T \sqrt{N} \gg m_{\rm p}$. This leads us to
conclude if that addition law was right, then our own sun would not exist.

Taken together we can clarify: In \cite{AmelinoCamelia:2011uk} is was shown that in the model considered
with a particular choice of coordinates, the soccer-ball problem is alleviated because colinear
momenta, and thus products of the average momentum itself, do not make a contribution. Unfortunately,
the argument presented in the paper does not suffice to solve the problem. 

\section*{Acknowledgements}

I thank Laurent Freidel, Jerzy Kowalski-Glikman, and Lee Smolin for helpful conversation.

\end{document}